\catcode`\@=11
\newif\if@fewtab\@fewtabtrue
{\count255=\time\divide\count255 by 60
\xdef\hourmin{\number\count255}
\multiply\count255 by-60\advance\count255 by\time
\xdef\hourmin{\hourmin:\ifnum\count255<10 0\fi\the\count255}}
\def\ps@draft{\let\@mkboth\@gobbletwo
    \def\@oddhead{}
    \def\@oddfoot
       {\hbox to 7 cm{$\scriptstyle Draft\ version:\ \draftdate$
       \hfil}\hskip -7cm\hfil\rm\thepage \hfil}
    \def\@evenhead{}\let\@evenfoot\@oddfoot}

\def\ceqno{\global\@fewtabfalse
    \ifcase\@eqcnt \def\@tempa{& & &}\or \def\@tempa{& &}
      \or \def\@tempa{&}
      \or\def\@tempa{}\fi\@tempa
{\rm(\theequation)}}

\def\aeqno#1{\global\@fewtabfalse
    \ifcase\@eqcnt \def\@tempa{& & &}\or \def\@tempa{& &}
      \or \def\@tempa{&}
      \or\def\@tempa{}\fi\@tempa
{\rm(\theequation,#1)}}

\def\label#1{\ifnum\draftcontrol=1
 \global\def\draftnote{$\scriptstyle #1$}\fi
 \@bsphack %
\def\draftnote{$\scriptstyle #1$}%
\if@filesw {\let\thepage\relax
   \def\protect{\noexpand\noexpand\noexpand}%
\xdef\@gtempa{\write\@auxout{\string
      \newlabel{#1}{{\@currentlabel}{\thepage}}}}}\@gtempa
   \if@nobreak \ifvmode\nobreak\fi\fi\fi
  \@esphack}

\def\alabel#1#2{\label{#1}\global\@fewtabfalse
    \ifcase\@eqcnt \def\@tempa{& & &}\or \def\@tempa{& &}
      \or \def\@tempa{&}
      \or\def\@tempa{}\fi\@tempa
{\hbox to 3cm{\phantom{\rm(\theequation,#2)}
\draftnote \hfil}\hskip -3cm {\rm(\theequation,#2)}}}

\def\clabel#1{\label{#1}\global\@fewtabfalse
    \ifcase\@eqcnt \def\@tempa{& & &}\or \def\@tempa{& &}
      \or \def\@tempa{&}
      \or\def\@tempa{}\fi\@tempa
{\hbox to 3cm{\phantom{\rm(\theequation)}
\draftnote \hfil}\hskip -3cm{\rm(\theequation)}}}

\def\eqnarray{\def\draftnote{{}}\global\@fewtabtrue
\stepcounter{equation}\let\@currentlabel=\theequation
\global\@eqnswtrue
\global\@eqcnt\z@\tabskip\@centering\let\\=\@eqncr
$$\halign to \displaywidth\bgroup\@eqnsel\hskip\@centering\@eqcnt\z@
  $\displaystyle\tabskip\z@{##}$&\global\@eqcnt\@ne
  \hskip 1\arraycolsep \hfil${##}$\hfil
  &\global\@eqcnt\tw@ \hskip 1\arraycolsep
$\displaystyle\tabskip\z@{##}$
\hfil  \tabskip\@centering&\global\@eqcnt\thr@@\llap{##}\tabskip\z@
\cr}

\def\endeqnarray{\@@eqncr\egroup
      \global\advance\c@equation\m@ne$$\global\@ignoretrue}

\def\@eqnnum{\hbox to 3cm{\phantom{\rm(\theequation)} \draftnote
                         \hfil}\hskip -3cm {\rm(\theequation)}}

\def\@@eqncr{\let\@tempa\relax
    \ifcase\@eqcnt \def\@tempa{& & &}\or \def\@tempa{& &}
      \or \def\@tempa{&}
      \or\def\@tempa{}
\fi\@tempa
\if@eqnsw
\if@fewtab\@eqnnum\fi
\stepcounter{equation}\fi\global
\@eqnswtrue\global\@eqcnt\z@\global\@fewtabtrue\cr}

\def\draftcite#1{\ifnum\draftcontrol=1#1\else{}\fi}

\def\@lbibitem[#1]#2{\item{}\hskip -3cm \hbox to 2cm
{\hfil$\scriptstyle\draftcite{#2}$}\hskip
1cm[\@biblabel{#1}]\if@filesw
     {\def\protect##1{\string ##1\space}\immediate
      \write\@auxout{\string\bibcite{#2}{#1}}}\fi\ignorespaces}

\def\@bibitem#1{\item\hskip -3cm \hbox to 2cm
{\hfil $\scriptstyle\draftcite{#1}$}\hskip 1cm
\if@filesw \immediate\write\@auxout
       {\string\bibcite{#1}{\the\value{\@listctr}}}\fi\ignorespaces}

\def\nsection#1{\section{#1}}%\setcounter{equation}{0}}

\def\nappendix#1{\vskip 1cm\no{\normalfont\Large\bfseries Appendix #1}\def\thesection{#1}}%\setcounter{equation}{0}}

\font\tendl=msbm10  scaled \magstep1%double line
\font\sevendl=msbm7 scaled \magstep1
\font\fivedl=msbm5 scaled \magstep1
\font\tengl=eufm10  scaled \magstep1% gothic letters
\font\sevengl=eufm7 scaled \magstep1
\font\fivegl=eufm5 scaled \magstep1

\newfam\dlfam \def\dl{\fam\dlfam\tendl} % \dl is double line
\textfont\dlfam=\tendl \scriptfont\dlfam=\sevendl
\scriptscriptfont\dlfam=\fivedl
\newfam\glfam \def\gl{\fam\glfam\tengl} % \gl is gothic letters
\textfont\glfam=\tengl \scriptfont\glfam=\sevengl
\scriptscriptfont\glfam=\fivegl

\def\draftdate{\number\month/\number\day/\number\year\ \ \ \hourmin }

\global\def\draftcontrol{0}
\catcode`\@=12
\def\tilde{\widetilde}

\def\theequation{\arabic{equation}} %%% no sections

\newcommand{\be}{\begin{eqnarray}}
\newcommand{\en}{\end{eqnarray}\vs 0.5 cm}

\newcommand{\no}{\noindent}
\newcommand{\vs}{\vskip}

\newcommand{\un}{\underline}
\newcommand{\NR}{{{\dl R}}}%letra doble raya en modo matematico
\newcommand{\NC}{{{\dl C}}}%letra doble raya en modo matematico
\newcommand{\NZ}{{{\dl Z}}}%letra doble raya en modo matematico
\newcommand{\qq}{\begin{eqnarray}}

\newcommand{\da}{\partial}
\newcommand{\ee}{{\rm e}}

\newcommand{\qqq}{\end{eqnarray}}

\newcommand{\tr}{\hbox{tr}}

\newcommand{\CA}{{\cal A}}

\newcommand{\CC}{{\cal C}}

\newcommand{\CP}{{\cal P}}

\newcommand{\s}{\hspace{0.05cm}}

\newcommand{\hf}{{_1\over^2}}

\newcommand{\lle}{\langle}
\newcommand{\rle}{\rangle}
\documentstyle[11pt]{article}
\newcommand{\bz}{{\bar{z}}}
\newcommand{\balpha}{\mbox{\boldmath$\alpha$}}
\begin{document}

\title{Chern-Simons theory and BCS superconductivity}
\author{\  Manuel Asorey, Fernando Falceto
\\ {\it Departamento de  F\'{\i}sica
Te\'orica, Univ. Zaragoza, Spain}
\\
 and 
\\ 
Germ\'an Sierra \\ 
{\it Instituto de Matem\'aticas y F\'{\i}sica Fundamental, CSIC, Spain}
}
\date{ }
\maketitle

%%% for draft versions, suppress in definitive version:
%\draft
%%
%%% suppress in definite version:
%\vskip 0.3cm

%%%
\vskip 1 cm

\begin{abstract}
\vskip 0.3cm
\noindent 
We study the relationship between the holomorphic unitary connection
of Chern-Simons theory with temporal Wilson lines 
and the Richardson's exact solution  of the reduced BCS Hamiltonian.
We derive the integrals of motion of the BCS model, their
eigenvalues and eigenvectors as a limiting case of the Chern-Simons theory. 
\end{abstract}
\vskip 2cm

\nsection{Introduction}
$\,$

 Chern-Simons theory has emerged in the last years
as a very useful  field 
theoretical model for the description
of interesting phenomena in Condensed Matter Physics. It plays a
leading role in the description of the quantum Hall effect \cite{JJ}. It has
been also advocated to be connected with hight $T_c$ superconductivity
 \cite{Letal}.
In this paper we will show that  it also appears to be closely related
to the Bardeen, Cooper and Schrieffer
(BCS) theory of superconductivity \cite{BCS}.

One of the corner stones in Condensed Matter Physics is 
the pairing model of Bardeen, Cooper and Schrieffer
(BCS) which explains the properties
of ``low $T_c$'' superconductors \cite{BCS} and several 
nuclei \cite{nuclear}.
The BCS model was solved in the Grand Canonical Ensemble
by a variational ansatz, which is asymptotically
exact in the number of fermions forming pairs. 
A reduced version of the BCS model, 
where the pairing interaction has
the same strength between all pairs, was solved exactly 
in 1963 by Richardson in the Canonical Ensemble, 
where the number of Cooper pairs is fixed \cite{R1,R1bis,RS}. 

Despite of the interest of this exact solution  
it escaped the attention of the condensed matter and nuclear physics
communities, until recently with 
the advent of ultra-small superconducting
grains \cite{RBT}, whose study is based on the BCS model
with a fixed number of pairs (for a review see \cite{vDR}). 

Partially motivated by Richardson's work, in 1976 Gaudin proposed
a family of commuting spin Hamiltonians whose exact diagonalization 
paralleled Richardson's solution \cite{G}. However a relation between
the two models was not clear at that time, despite of the
fact that  in the limit
where the BCS coupling constant $g$ goes to infinity, the Richardson's
solution turns into the Gaudin's one. 

A first step to understand 
the relation between the Gaudin's spin Hamiltonians and the reduced 
BCS Hamiltonian came with  
the work of Cambiaggio, Rivas and Saraceno (CRS) who
proved the integrability of the reduced BCS model in terms
of a family of commuting operators, which are  
nothing but the Gaudin's Hamiltonians plus a perturbation
proportional to the inverse of the BCS coupling constant $g$ \cite{CRS}. 
CRS also constructed the BCS Hamiltonian as a linear combination
of the perturbed Gaudin's operators.  
Unfortunately these authors were unaware of Richardson's
and Gaudin's works and consequently  they did not obtained 
the eigenvalues of the BCS conserved quantities. 

The eigenvalues of the CRS operators were given in the reference
\cite{S}, using Conformal Field Theory (CFT) methods, and generalized
to other models in \cite{DS,ALO,DES}. The aim of reference \cite{S}
was to place the Richardson's exact solution and its integrability 
in the framework of CFT, relating it to 
the Wess-Zumino-Witten models with Kac-Moody algebra $SU(2)_k$. 
Using CFT methods it was shown that the Richardson's wave function 
are obtained as conformal blocks of a perturbed $SU(2)_k$--WZW model when
the level $k$ goes to a critical value $-2$ \cite{S}.
 
The CFT interpretation of the BCS model 
turned out to be closely related 
to the work of Babujian \cite{B}, who in 1993,
used the so called off-shell algebraic Bethe ansatz, to 
re-derived  Gaudin's exact solution \cite{B,BF}. 
Babujian also observed  that 
the  Gaudin's eigenstates can be used to build the conformal blocks
of the WZW models.

The similarity between the works
\cite{S} and \cite{B,BF} suggested 
that the Richardson's solution of the reduced 
BCS model should also be derivable using the off-shell Bethe ansatz
method. This was done  by Amico, Falci and Fazio \cite{AFF} and
later on clarified in references \cite{ZLMG,vDP,S2},  
where the BCS coupling
constant parametrizes a boundary operator that appears in the transfer 
matrix of the inhomogenous vertex model, whose semi-classical
limit gives rise to the CRS conserved quantities.

The old results by Richardson and Gaudin's, concerning
the norm of the eigenstates and the occupation numbers
\cite{R2,Gbis},  
has been generalized to other operators in references
\cite{AO,ZLMG} using the ``determinantal'' techniques
developed in \cite{K,Sk}.

The outcome of all these works has been to clarify the
integrability and relationship between the reduced BCS and Gaudin's
model at least at a formal level. 
There remains however the question concerning 
the ``geometrical'' or field theoretical 
origin of the integrability of BCS. 
For the Gaudin's Hamiltonians this is given by
the underlying WZW model, and ultimately by the Chern-Simons
theory \cite{W}. 
The BCS model is related to some sort 
of chiral perturbation of the WZW model, characterized by modified
Kniznik-Zamolodchikov equations \cite{S,ZLMG,S2}. 

In this paper we shall
show that the field theoretical origin of BCS  
can be traced back to a $SU(2)$ Chern-Simons (CS) theory interaction with
a  one-dimensional distribution of coloured matter which breaks both
gauge and conformal invariance. This connection is quite remarkable because
Chern-Simons theory has  been advocated to be mainly connected with
effective descriptions of fractional quantum Hall effect 
and high T$_c$ superconductivity, 
but never with standard BCS superconductivity. The Chern-Simons theory is
not defined in the physical space, which might be
two or three-dimensional, but rather in the complex energy plane 
which is always two-dimensional. This explains why 
this field theoretical connection of BCS theory remained
unveiled for so long time.
The connection of Chern-Simons theory with BCS model can be understood in
a more general framework when we consider a scaling limit
of the twisted Chern-Simons theory defined on a torus, that is, 
the twisted elliptic Chern-Simons theory.  
On a torus the KZ equations \cite{KZ} are replaced
by the Knizhnik-Zamolodchikov-Bernard (KZB) equations \cite{Bernard},
which depend on the  coordinates $z_n$ 
of the punctures, the moduli of the torus
$\tau$ and a set of parameters $u_j$ characterizing 
the toroidal flat
gauge connections. The later parameters
$u_j$ define the twisted boundary conditions for the
WZW fields on the torus. 

The main result of this paper
is to show that, for a generic simple
simply connected, compact Lie group $G$, the Richardson
equations, the CRS conserved quantities and their eigenvalues 
arise from the KZB connection and their associated 
horizontal sections. This will be done in a limit 
where the torus degenerates into the cylinder and then 
into the complex plane. In this limiting procedure the generalized
BCS coupling constants appear as conjugate variables
of the parameters $u_j$, when this parameters go to infinity.
This gives the $G$-based BCS models a suggestive
geometrical and group theoretical meaning.

The organization of the paper is as follows.
In section 2 we briefly  review the reduced
BCS model and its exact solution. In section 3
we consider a perturbation of the CS model defined
on the plane, we derive the KZ equations 
and study their connection with the CRS
conserved quantities and the BCS model. 
In section 4 we introduce the 
KZB connection related to the CS theory on the
torus. In section 5 we derive the Bethe ansatz
of the integrable Hamiltonians introduced in 
sections 3 and 4. Finally we state our conclusions.
We have also included an appendix with 
particular examples of the general equations
contained in the main text.

\nsection{ Review of the exact solution of 
the BCS model}

The reduced BCS model is defined by the Hamiltonian 
\cite{BCS,vDR,S}
\begin{eqnarray}
H_{BCS} = \sum_{n, \sigma= \pm} 
\varepsilon_{n\sigma} c_{n \sigma}^\dagger c_{n \sigma}^{}
  -g d \sum_{n, n'}  c_{n +}^\dagger c_{n -}^\dagger 
c_{n' -}^{} c_{n' +}^{} \; 
\label{1}
\end{eqnarray}
\noindent where $c_{n,\pm}$ (resp. $c^\dagger_{n,\pm}$)
is an electron  
destruction (resp. creation) operator   
in the time-reversed states $|n, \pm \rangle$
with energies $\varepsilon_n$, $d$ is the  mean level spacing  and
$g$ is the BCS dimensionless coupling constant.
The sums in (\ref{1}) run over a set of $N$ doubly
degenerate energy levels $\varepsilon_n (n=1,\dots, N)$. 
We shall assume that the energy levels  are all
distinct, i.e. $\varepsilon_m \neq \varepsilon_n $ for $m \neq n$.  
The Hamiltonian (\ref{1}) is a simplified 
version of the reduced BCS Hamiltonian where all couplings
have been  set equal to a single one,  namely $g$. 
This is the model that is commonly used to describe
ultrasmall grains and accounts for the scattering 
of pairs of electrons between discrete energy levels
that come in time-reversed states. 
Hereafter  we shall refer to  
(\ref{1}) simply as the BCS Hamiltonian. 

Richardson had long ago solved this model exactly for 
an arbitrary set of levels, $\varepsilon_n$ \cite{R1,R1bis,RS}. 
To simplify matters,  we shall assume that
there are not singly occupied electronic levels. 
As can be seen from (\ref{1}), these levels decouple
from the rest of the  system;  
they are said to be blocked,  
contributing only  with their energy  
$\varepsilon_n$ to the total energy $\cal E$. 
The above simplification implies that every energy level
$n$ is either empty (i.e. $| {\rm vac}\rangle$),
or occupied by a pair of electrons (i.e.
$c^\dagger_{n,+} c^\dagger_{n,-} |{\rm vac}\rangle$). 
Denote the total number of electrons pairs by 
$M$. Then  of course
$M \leq N$. The most studied case
in the literature corresponds to the half-filled situation,  
where the number of electrons,  $N_e= 2M$, is  equal to the number of levels
$N$ \cite{vDR}. 
In the absence of interaction (i.e. $g =0$),  
all the  pairs occupy the lowest  energy levels forming a  
Fermi sea. The pairing interaction promotes
the  pairs to  higher  energies  and eventually,
for large values of $M$, all the levels are pair correlated,
giving rise to superconductivity \cite{BCS}.

In order to describe Richardson's solution one
defines the hard-core boson operators 
\qq
b_n = c_{n,-} c_{n,+}, \;\; b_n^\dagger= 
c^\dagger_{n,+} c^\dagger_{n,-} , \;\; N_n = b^\dagger_n b_n 
\label{2}
\qqq
\noindent which satisfy the commutation relations,
\qq
[ b_n^{}, b_{n'}^\dagger ] = \delta_{n,n'} \; (1 - 2 N_n) 
\label{3}
\qqq

The Hamiltonian (\ref{1}) can then be written as
\qq
H_{BCS} = \sum_{n} 2 \varepsilon_n b^\dagger_n b_n^{} - g \, 
\sum_{n,n'} \; b_n^\dagger b_{n^{\prime}}^{} \; ,
\label{4}
\qqq
\noindent where we have set $d=1$ (i.e. all the energies
are measured in units of $d$). 
Richardson showed that the eigenstates of this Hamiltonian 
with $M$ pairs have the (unnormalized) product form 
\cite{R1,R1bis,RS} 
\begin{eqnarray}
|M \rangle_R = \prod_{j = 1}^M B_j |{\rm vac} \rangle, \;\;
B_j = \sum_{n=1}^N \frac{1}{2 \varepsilon_n - E_j} 
\; b^\dagger_n
\label{5} 
\end{eqnarray}
\noindent 
where the  parameters $E_j$ ($j = 1, \dots ,M$) are, 
in general, complex solutions of the $M$ coupled algebraic
equations 
\qq
\frac{1}{g } + \sum_{j'=1 (\neq j)}^{M} \frac{2}{ E_{j'} - E_j} 
= \sum_{n=1}^N \frac{1}{2 \varepsilon_n - E_j} \; , 
\label{6}
\qqq
\noindent 
which are a sort of Bethe ansatz equations for this
problem. 
The energy of these states is given by the sum of the
auxiliary parameters $E_j$, i.e.
\qq
{\cal E} (M) = \sum_{j = 1}^{M} E_j
\label{7}
\qqq
\noindent 
The ground state  of $H_{BCS}$ is given by the solution
of eqs.(\ref{6}) which gives the lowest value of ${\cal E}(M)$. 
The (normalized) states (\ref{5}) can also be written  
as  \cite{R2}
\qq
|M \rangle_R = \frac{C}{ \sqrt{M!} } \sum_{n_{_1}, \dots , n_{_M}=1}^N
\psi^R(n_{_1}, \dots,  n_{_M})\ b^\dagger_{n_{_1}} \cdots\ b^\dagger_{n_{_M}}
|{\rm vac} \rangle 
\label{8}
\qqq
\noindent where 
the sum excludes double occupancy of pair states
($n_i\neq n_j$ for $i\neq j$) 
and the wave function $\psi$ takes the form
\qq
\psi^R(n_{_1}, \dots, n_{_M}) = \sum_{\CP\in S_M}
 \prod_{k=1}^{M} \frac{1}{ 2 \varepsilon_{n_k} 
- E_{{\CP}k} }  
\label{9} 
\qqq
\noindent The  sum in (\ref{9})  runs  over the set
of permutations $S_M$, of $1, \cdots, M$. 
The constant  $C$ in (\ref{8}) guarantees the normalization 
of the state \cite{R2} (i.e. $_R\langle N|N\rangle_R =1$).

A well known fact about the BCS Hamiltonian is that
it is  equivalent to that of a XY model with long
range couplings and a ``position dependent'' 
magnetic field proportional to $\varepsilon_n$.
To see this let us represent the hard-core boson operators
(\ref{2})  in terms of the Pauli 
matrices as follows,
\qq
b_n = \sigma_n^+, \; b_n^\dagger = \sigma_n^- ,\; 
N_n = \frac{1}{2} (1 - \sigma_z) 
\label{10}
\qqq
\noindent 
in which case the Hamiltonian (\ref{4}) becomes
\qq
\begin{array}{l}
\displaystyle
 H_{BCS} = H_{XY} + \sum_{n=1}^N \varepsilon_n +
 g (N/2 -M)   \\
\displaystyle
 H_{XY}= - \sum_{n=1}^N 2 \varepsilon_n \tau^0_n - \frac{g}{2}
( T^+ \; T^- + T^- \; T^+ )  
\end{array}
\label{11}
\qqq
\noindent where the matrices 
\qq
\begin{array}{l}
\displaystyle
 T^a = \sum_{n=1}^N  \tau^a_n \;\; (a = 0, +, -)  
 \\
\\
\displaystyle
 \tau^0_n = {\textstyle\frac{1}{2}} \sigma^z_n, \;\; \tau^+_n = \sigma^+_n ,
\;\; \tau^-_n = \sigma^-_n  
\end{array} 
\label{12}
\qqq
\noindent satisfy the $SU(2)$ algebra,
\qq
\begin{array}{l}
\displaystyle
 [ T^a, T^b] = f^{a b}_c T^c   
\\
\\
\displaystyle
 f^{+0}_+ = f^{0-}_- = -1, \; f^{+-}_0 = 2 
\end{array}
\label{13}
\qqq
\noindent whose Casimir is given by 
\qq
{\bf T} \cdot {\bf T} = T^0 T^0 +
\frac{1}{2} ( T^+ T^- + T^- T^+) 
\label{14}
\qqq

This spin representation of the BCS model
is the appropriate one to 
study its integrability, as was shown 
by Cambiaggio, Rivas and Saraceno
by explicitly constructing a set
of operators, $R_n \;\;(n=1, \dots,N)$, which commute 
with $H_{BCS}$ and moreover allow a reconstruction
of the BCS Hamiltonian which is given by
$H_{BCS} = g \sum_n 2 \epsilon_n \; R_n + {\rm ctes}$.
In the next sections we shall generalized this 
construction to arbitrary Lie groups and representations.

\nsection{Chern-Simons Theory and BCS Hamiltonian.}
Let $G$ be a simple, simply connected, 
compact Lie group. 
The Chern-Simons theory of $N$ interacting  
heavy coloured particles (temporal Wilson line insertions) located at 
points $z_1,z_2,\dots,z_N$ of the 2D plane $\NC$
is defined by the action \cite{GL}
\qq
\begin{array}{l}
\displaystyle
S(A,g_1,\dots,g_N) = {k\over 4 \pi} \int \! \ {\rm tr}\left(A
\wedge d A +
{2\over 3} A\wedge A\wedge A\right) \\
\displaystyle 
\phantom{S(A,g) = }+\sum_{n=1}^{N}\int_{-\infty}^{\infty} \  dt \
\ \langle
 \Lambda_n,g_n^{-1}(t)A_0(z_n,t) g_n(t)+ g_n^{-1}(t){d\over dt}g_n(t)\rangle
\\
\end{array}
\label{csa}
\qqq
where $A$ is a  $G$-gauge field, $g_n$ a $G$-valued chiral field describing
the dynamics of  the coloured particle located at the insertion point $z_n$
 and  
$\Lambda_n$ the  highest weight of its $G$-colour representation. 
 $\lle\cdot \,,\cdot\rle$ denotes the Killing scalar product
in the Lie algebra ${\gl g}$ of $G$ normalized 
so that long roots  have length $2$, we use this scalar product 
to identify 
${\gl g}$ and its dual ${\gl g^\ast}$.
The quantization of the theory (\ref{csa}) yields quantum states 
which have to satisfy two constraints. In the Bargmann representation
they are holomorphic functionals $\Psi(A_\bz)$ of the 
$(0,1)$ component of the gauge field $A_\bz= \hf(A_1+i A_2)$ %\cite{EMSS}
and take values in the tensor product of the representation spaces 
$\otimes V_{\Lambda_n}$ of the insertions. The remaining constraint 
is Gauss' law
\qq
D^{ab}_{\bz}{\delta\over\delta A^b_\bz(z)}\Psi(A_\bz)=
-{k\over\pi}\da_z A^a_\bz \Psi(A_\bz)+\sum_{n=1}^N  t^a_{(n)} 
\delta(z-z_n) \Psi(A_\bz),
\label{hgl}
\qqq
where $t^a$ is an orthonormal basis of the Lie
algebra ${\gl g}$ of $G$ and 
 $ t^a_{(n)}$
denotes the operator representation on $V_{\Lambda_n}$ of 
  $t^a\in {\gl g}$.
The law (\ref{hgl}) governs the dependence of the quantum states 
$\Psi$ under a 
$G^\NC$--valued complex gauge transformation $g$, i.e. 
\qq
\Psi(g^{-1}
A_\bz g+ g^{-1}\da_\bz g)=\ee^{k S_{WZW}(A_\bz,g)} \prod_{n=1}^N g(z_n)^{-1}_{(n)}
\Psi(A_\bz)
\label{jgl}
\qqq where
\qq
\begin{array}{l}
\displaystyle
S_{WZW}(A_\bz,g)= -{i\over {4\pi}}
\int  d z d \bz\  \tr\, g^{-1}\, (\partial_z\, g)g^{-1}{\partial}_\bz\,g
-{i\over {12\pi}}\int d^{-1}\tr\,(g^{-1}d g)^3 \\
\displaystyle
\phantom{S_{WZW}(A,g)=} + {i\over 2 \pi}
\int  d z d \bz\  \tr\, A_\bz \, g\da_z g^{-1}
\end{array}
\nonumber
\qqq
is the Wess-Zumino-Witten action.
In equation (\ref{jgl}) $ g_{(n)}$
denotes the operator representation on $V_{\Lambda_n}$ of 
the element $g$ of $G^\NC$.
The global implementation of this dependence
requires that the coupling constant $k$ is an integer.
Since any gauge field $A$ can be rewritten as  $A= g^{-1}\da_\bz g$ 
in terms of a complex gauge transformation $g$, all
physical states are determined by
its value at the trivial gauge field configuration $A_\bz=0$, i.e. 
$\Psi(0)$. On this configuration, the infinitesimal dependence of the physical
states on the gauge field can be obtained \cite{LR} from Gauss' law (\ref{hgl})
\qq
\pi {\delta\over\delta A^a_\bz(z)}\Psi(A_\bz)\Big|_{A=0}
= \sum_{n=1}^N { t^a_{(n)} 
\over z-z_n} \Psi(0)
\label{fvn}
\qqq
and
\qq
\begin{array}{l}
\displaystyle
\pi^2{\delta^2\over\delta A^a_\bz(z)\delta A^a_\bz(w)}\Psi(A_\bz)\Big|_{A=0}=
{k \dim G\over (z-w)^2}\Psi(0) +
\sum_{n=1}^N { t^a_{(n)} t^a_{(n)}\over (w-z_n)(z-z_n)}\Psi(0)
\\
\displaystyle
\phantom{
\pi^2{\delta^2\over\delta A^a_\bz(z)\delta A^a_\bz(w)}\Psi(A_\bz)\Big|_{A=0}=}
+\sum_{n\neq m} { t^a_{(n)} t^a_{(m)} 
\over (w-z_n)(z-z_m)} \Psi(0),
\end{array}
\label{svn}
\qqq
where the sum in the repeated index $a$ is implicitly assumed and 
%$C_v^{(n)}$ is the value of the quadratic Casimir in the representation
%$\Lambda_n$ and 
$1/(z-w)$ is the Green function of the operator 
${1\over \pi}\da_\bz$. From these expressions and the Sugawara 
form of the stress tensor 
\footnote{
The metric dependence of the theory is induced both by
the Bargmann normalization of quantum
states ($k$) and the removal of spurious gauge degrees of freedom ($h^\vee$)}
\qq
T_{zz}={\pi^2\over 2(k + h^\vee)}:{\delta^2\over\delta A^a_\bz(z)\delta A^a_\bz(z)}:= {\pi^2\over 
2(k + h^\vee)}\lim_{w\to z}
\Big[{\delta^2\over\delta A^a_\bz(z)\delta A^a_\bz(w)}-
{k \dim G\over (z-w)^2}\Big]
\label{stress}
\qqq
it is straightforward to show that the dependence on the
translation of the insertion at $z_n$
is governed by the Knizhnik-Zamolodchikov equation \cite{KZ}
\qq
\da_{z_n}\Psi(0) +
 {1\over \kappa}\sum_{m\neq n} { t^a_{(n)} t^a_{(m)} 
\over z_m-z_n} \Psi(0)=0,
\label{pKZ}
\qqq
where $\kappa=k + h^\vee$ and $h^\vee$ is the dual Coxeter number.
In the same way the dilation transformation of the plane implies that
\qq
\sum_{n=1}^N z_n \da_{z_n}
\Psi(0) + {1\over\kappa} H_G \Psi(0)=0
\qqq
where the operator
\qq
H_G=-{1\over2}\sum_{n\neq m} { t^a_{(n)} t^a_{(m)}}
\label{gd}
\qqq
in the second term of the equation 
is essentially the coupling term in the 
BCS Hamiltonian (eq ({\ref{11})). 
This suggest that a possible slight modification of the Chern-Simons
theory can trace back the geometric origin of the BCS Hamiltonian.

Let us consider the perturbation of Chern-Simons theory by a one-dimensional
distribution of coloured charges. The effect of this perturbation is encoded
by a new interaction term of the form  
\qq
S_{\rm \xi}(A)=-{1\over 2 \pi i}\int_{-\infty}^{\infty}dt \, \oint_\CC dz \, 
{\rm }  \langle \xi\, ,A_0(z,t) \rangle
\nonumber
\qqq
where $\xi$ is any element of the complexified Lie algebra 
${\gl g}^\NC$, $\CC$  is a non-self-intersecting closed curve  
of $\NC$ which encloses all point-like 
insertions $z_n$ of heavy coloured particles
and where a one-dimensional $\xi$-coloured charge distribution
is localized. Because of its $\xi$ dependence this perturbation is not
gauge invariant under gauge transformations not completely 
included in the normalizer of $\xi$.
On the other hand, conformal invariance is
broken by the special choice of the closed curve 
 $\CC$. However, the full Chern-Simons theory defined by $S(A,g)+
S_{\rm \xi}(A)$ can be quantized in a similar way and
the corresponding quantum states are holomorphic  
functionals $\Psi(A_\bz)$ taking values in the same vector space
$\otimes V_{\Lambda_n}$ because the new  insertions have 
one-dimensional representations. The  essential difference is
encoded by the Gauss' law which now reads
\qq
\begin{array}{l}
\displaystyle
D^{ab}_{\bz}{\delta\over\delta A^b_\bz(z)}\Psi(A_\bz)=
-{k\over\pi}\da_\bz A^a_\bz \Psi(A_\bz)+\sum_{n=1}^N  t^a_{(n)} 
\delta(z-z_n) \Psi(A_\bz) \\ \displaystyle
\phantom{D_{\bz}{\delta\over\delta A_\bz(z)}\Psi(A_\bz)=a}
-{1\over 2 \pi i}\oint_\CC dz'
\delta(z-z')\langle \xi,t^a\rangle \Psi(A_\bz).
\end{array}
\label{hmgl}
\qqq
This implies that the variation (\ref{svn}) now becomes 
\qq
\begin{array}{l}
\displaystyle
\pi^2{\delta^2\over\delta A^a_\bz(z)\delta A^a_\bz(w)}\Psi(A_\bz)\Big|_{A=0}=
{k \dim G\over (z-w)^2}\Psi(0)+
\sum_{n=1}^N {t^a_{(n)} t^a_{(n)}\over (w-z_n)(z-z_n)}\Psi(0) 
\\ \displaystyle
\phantom{adf} +\sum_{n\neq m} { t^a_{(n)} t^a_{(m)} 
\over (w-z_n)(z-z_m)} \Psi(0) -
\sum_{n=1}^N \left({ \xi_{(n)}
\over z-z_n}+ { \xi_{(n)}
\over w-z_n}\right)\Psi(0)+ \langle \xi,\xi\rangle \Psi(0)\\ 
\end{array}
\label{msvn}
\qqq
where $ \xi_{(n)}$ denotes the action of $\xi$ on the
representation $\Lambda_n$ of the coloured particle sitting at $z_n$.
Using the Sugawara form of the stress tensor (\ref{stress})
it is easy to show that the physical states must 
satisfy a modified Knizhnik-Zamolodchikov equation \cite{KZ}
\qq
\da_{z_n}\Psi(0) +{1\over \kappa} \xi_{(n)}\Psi(0) +
 {1\over \kappa}\sum_{m\neq n} { t^a_{(n)} t^a_{(m)} 
\over z_m-z_n} \Psi(0)=0.
\label{mpKZ}
\qqq
In a similar way the dilation transformation now yields the equation
\qq
\sum_{n=1}^N z_n \da_{z_n} 
\Psi(0) + {1\over\kappa} H\Psi(0) = 0
\qqq
and, what is remarkable, in the case $G=SU(2)$ the operator
\qq
H=
- {1\over 2} \sum_{n\neq m} { t^a_{(n)} t^a_{(m)}}
+\sum_{n=1}^N z_n \xi_{(n)}
\label{BCS}
\qqq
 does coincide up to a factor $1/g$ and additive constant
with the reduced BCS Hamiltonian (\ref{11}).

The solvability of the reduced BCS Hamiltonian worked out by Richardson can
now be understood in this geometrical
approach from the existence of a complete set of
operators 
\qq
R_n= 
\sum_{m\neq n} { t^a_{(n)} t^a_{(m)}\over z_m -z_n}+\xi_{(n)}
\label{rchs}
\qqq
commuting with $H$, $[R_n,H]=0$, as first pointed 
out in Ref. \cite{CRS}. The operators  $R_n$ also commute each other
($[R_n,R_m]=0$) and provide a complete set of
constants of motion.
This remarkable feature  generalizes  the well known
integrability of the Hamiltonian (\ref{gd}) which commutes with
the Gaudin Hamiltonians. The only difference is due to the presence of
the $\xi$ term in (\ref{rchs}). 
The commutativity of the Hamiltonians $R_n$ also means that the connection
(Knizhnik-Zamoldchikov connection)
defined by them in the bundle of  Chern-Simons states over the
space of insertions $\NR^N$ is flat. In fact, this flatness property of
the KZ connection is not only verified on the bundle of CS states 
but in a more general bundle of states where the extra condition
of global colour invariance imposed by
Gauss' law is not required. This colour neutrality 
condition comes from the integration
of Gauss' law equation (\ref{hmgl}) and establishes 
that CS states  can only take values
on the subspace $V^0\subset\otimes V_{\Lambda_n}$
of invariants under the diagonal action of the Cartan subgroup of $G$
(here and below we take $\xi$ in the Cartan subalgebra of G)
and as we shall see later it will correspond to the half-filling regime
of BCS superconductivity. In fact, this restriction can
be ignored if we include an extra charge insertion $z_0$ at very 
large distance of all the other insertions with the only purpose of
neutralizing their colour charge.
In this way, the whole BCS theory emerges from CS 
in the singular limit $\kappa\to 0$
as we shall see in next section.
This relation existing between Chern-Simons
theory and the BCS theory can be even better understood 
from a elliptic point of view.

\nsection{Knizhnik-Zamoldchikov-Bernard connection.}

Let us consider now the  Chern-Simons theory on a compact 
two dimensional space $T^2$ with genus one. This theory 
was studied in \cite{FG1,FG2}}. Let us summarize the main results.
To quantize the theory one has to pick up a complex structure in the torus
$\NC/(\NZ+\tau\NZ)$ with $\tau_2\equiv {\rm Im}\tau>0$ and a local systems
of coordinates $z_n$ for the n$^{th}$ 
Wilson line insertions. Then, quantization
proceeds in a similar way as for the plane or the sphere.
In  Bargmann's
quantization  the states are holomorphic functionals of the $(0,1)$
part of the gauge 1-form with values in the tensor product
of the representation spaces $\otimes V_{\Lambda_n}$.
The states are subject to the Gauss' law that once integrated
tells us how the states transform under complex gauge
transformation. 
Every (generic) gauge field can then be gauged out
to a flat one of the form $A_u=\pi ud\bar z/\tau_2$ with $u\in{\gl h}^{\NC}$,
the (complexified) Cartan subalgebra of $G$, and 
the states are effectively described in terms multidimensional 
theta functions $\theta(u)$
with values in the subspace $V^0\subset\otimes V_{\Lambda_n}$.
The states are also  subject to extra regularity conditions. 
See \cite{FG1,FG2,EFK} for details.

The dependence of the quantum states on the positions of the 
point-like insertions can be derived in a similar manner as in the planar case.
However, in the elliptic case 
Gauss' law (\ref{hgl}) does not provides in a simple manner
the infinitesimal variation of quantum states with respect to arbitrary
gauge fields. In fact, the required Green function of the operator
$D_\bz=\da_\bz+\pi u/\tau_2$ does depend on $u$. The trick 
is to remove the $u$ dependence
from the differential operator $D_\bz=\da_\bz+\pi u/\tau_2$ by means
of a singular 
gauge transformation
$$\Psi(u)=\ee^{\pi k \langle u\, , u\rangle
/2\tau_2}\prod_{n=1}^N
\ee^{-\pi u_{(n)} (\bz_n-z_n)/\tau_2}\gamma(u).$$
The $u$ dependence is traded in this manner
by a non-trivial twisted boundary condition
for Chern-Simons states and the  Green functions of the operator $\da_\bz$.
We further extract a normalization 
factor accounting for the volume of the complex
gauge orbits
$$\Pi(u)= \ee^{\pi i \tau{\rm dim}G /12}\prod_{\alpha>0}\left(
\ee^{\pi i\lle u,\alpha\rle}-\ee^{-\pi i\lle u,\alpha\rle}\right)
\prod_{l=1}^\infty\left(1-\ee^{2\pi i \tau l}\right)^{{\rm rank} G} 
\prod_{\alpha}\left(1-\ee^{2\pi i \tau l+2\pi i \lle u,\alpha\rle }\right)
$$
with the product in $\alpha$ (resp. $\alpha>0$) extending to the set of roots
(resp. positive roots) of $G$.   
If we then define $\theta$ by 
$$\gamma(u)= \Pi(u) \theta(u),$$
the dependence of the CS quantum states on the
coordinates of the point-like charged insertions  and
on the elliptic modular parameter $\tau$ \cite{ADW} 
can be expressed in a simple
geometrical meaningful form: 
Chern-Simons states describe horizontal sections
$$\nabla_\CA \theta(u)=0$$ with
respect to the flat Knizhnik-Zamolodchikov-Bernard (KZB) connection 
$\CA$ defined by 
\qq
\CA_{\bar\tau}\s\s=0,\quad
\CA_{\bar z_n}= 0,\quad
\CA_{\tau}\s\s=\s{_1\over^\kappa} H_0(\tau,\un{z})\s,\quad
\CA_{z_n}=\s{_1\over^\kappa}\s H_n(\tau,\un{z})\s.
\label{KZB}
\qqq
with
\qq
\begin{array}{l}
\displaystyle
H_0(\tau,\un{z})={i\over{4\pi}}\Delta_u+
{i\over{8\pi}}\sum\limits_{n,m=1}^N\bigg(
-2\sum\limits_\alpha \da_x
\sigma_{\lle u,\alpha\rle}(z_{n}-z_{m})
\s\s(e_\alpha)_{_{(n)}}(e_{-\alpha})_{_{(m)}}\\
\phantom{H_0(\tau,\un{z})={i\over{4\pi}}\Delta_u+
{i\over{8\pi}}\sum\limits_{n,m=1}^N\bigg(}
\displaystyle
+ \sum\limits_{j=1}^r(\rho(z_{n}-z_{m})^2+
\rho'(z_{n}-z_{m}))\s\s h^j_{_{(n)}}h^j_{_{(m)}}
\bigg)\s,
\end{array}
\label{H0}
\qqq
\qq
\begin{array}{l}
\displaystyle
H_n(\tau,\un{z})=\sum\limits_{j=1}^rh^j_{_{(n)}}\da_{u^j}\s-
\s\sum\limits_{m\not=n}\bigg(\sum\limits_{\alpha}
\sigma_{\lle u,\alpha\rle}(z_{n}-z_{m})\s\s
(e_\alpha)_{_{(n)}}(e_{-\alpha})_{_{(m)}}\\
\phantom{H_0(\tau,\un{z})={i\over{4\pi}}\Delta_u+
{i\over{8\pi}}\sum\limits_{n,m=1}^N ssss}
\displaystyle
+ \sum\limits_{j=1}^r
\rho(z_{n}-z_{m})h^j_{_{(n)}}h^j_{_{(m)}}
\bigg)
\end{array}
\label{Hn}
\qqq
on the vector bundle of Chern-Simons states
over the moduli of punctured elliptic curves. In equations 
(\ref{H0})(\ref{Hn})
$\rho$ and $\sigma_x$ denote the Green functions of 
${1\over\pi}\partial_{\bar z}$
that can be expressed in terms of the Jacobi theta function 
$$\vartheta_1(z)\s=\s\sum_l(-1)^l
\ee^{\s\pi i(l+1/2)^2\tau+2\pi i z(l+1/2)}$$  
by
$$\rho(z)={\vartheta'_1(z)\over\vartheta_1(z)}$$  
and 
$$\sigma_x(z)={\vartheta'(0)\vartheta(x-z)\over
\vartheta(x)\vartheta(z)},$$
and we use the notation of Refs. \cite{EK,FW,FV}:  
$e_\alpha$ represents the step 
generator associated to root $\alpha$ of ${\gl g}$ and
$\{h_i, i=1,\dots,r\}$ is an 
orthonormal 
basis of the Cartan subalgebra, so that $u=\sum_j u_j h_j$.

%%%%%%%%%%

A natural hermitian structure in this bundle was introduced in Ref. 
\cite{FG3} by means of
the scalar product \`a la Bargmann, 
%This product was shown to be
%finite in many particular cases and its finitness was conjectured to 
%hold for all Chern-Simons states \cite{FG3}.  
in that paper it was also shown that KZB connection is the unique
$(1,0)$ unitary connection in this bundle. 

%%%%%%%%%%%%%%
It is easy to check that the KZB connection leave $V^0$ invariant
and restricted to $V^0$ is flat for any value of $\kappa$. 
This implies, in particular, that Hamiltonians $H_n$ 
restricted to $V^0$ commute for $n=0,1,\dots,N$ and determine then
a quantum integrable system that admits
a family of eigenvectors using the appropriate Bethe ansatz \cite{FG3}.
Below we shall discuss the solutions to the problem of eigenvalues.

Notice that in this case the equation corresponding to
a global dilation is more involved that in planar case which makes
the connection with a Hamiltonian of BCS type less transparent.
However, we shall show that in the limit $\tau\to i\infty$
where the torus degenerates into the 
cylinder we shall recover a connection with
the BCS theory. In particular, after rescaling $\tau\mapsto\eta\tau$ 
and taking the limit 
$\eta\to\infty$ we obtain the new family of commuting Hamiltonians,
$\tilde H_0$ and $\tilde H_n$ that read:
\qq
\tilde H_0(\un{z})&=&{i\over{4\pi}}\Delta_u\s-
{\pi i\over4}
\sum\limits_\alpha
{{\bf e}_\alpha{\bf e}_{-\alpha}\over \sin^2(\pi \lle u,\alpha\rle)}
\s,
\label{tauinfH0}
\qqq
\qq
\begin{array}{l}
\displaystyle
\tilde H_n(\un{z})=\sum\limits_{j=1}^rh^j_{_{(n)}}\da_{u^j}- 
\pi \sum\limits_{m\not=n} \cot(\pi(z_{n}-z_{m}))t^a_{_{(n)}} t^a_{_{(m)}} \\
\phantom{ H_n(\un{z})=\sum\limits_{j=1}^rh^j_{_{(n)}}\da_{u^j} }
\displaystyle
 + \pi\sum\limits_{m\not=n}\sum\limits_{\alpha}
\cot(\pi \lle u,\alpha\rle)
(e_\alpha)_{_{(n)}} (e_{-\alpha})_{_{(m)}}
\end{array}
\label{tauinfHn}
\qqq
with ${\bf e}_\alpha=\sum_n (e_\alpha)_{_{(n)}}$.

Next we take the limit from the cylinder to the plane by
sending variables $z_n$ and $u$ to the origin, we rescale 
$z_n\mapsto\zeta z_n$ and $u\mapsto\zeta u$, 
$\tilde R_0=\zeta^2\tilde H_0$, $\tilde R_n=\zeta\tilde H_n$. 
In the limit of $\zeta\to 0$ the new commuting Hamiltonians 
are:
\qq
\tilde R_0(\un{z})&=&{i\over{4\pi}}\Delta_u-
{i\over4\pi}
\sum\limits_\alpha
{{\bf e}_\alpha{\bf e}_{-\alpha}\over \lle u,\alpha\rle^2}
\s,\alabel{tildeR0}a\cr
\tilde R_n(\un{z})&=&
\sum\limits_{m\not=n}{t^a_{_{(m)}}t^a_{_{(n)}}\over
z_{m}-z_{n}}+\sum\limits_{j=1}^rh^j_{_{(n)}}\da_{u^j}-
\s\sum\limits_{m\not=n}\sum\limits_{\alpha}
{(e_\alpha)_{_{(m)}}(e_{-\alpha})_{_{(n)}}\over \lle u,\alpha\rle}
\alabel{tildeRn}b\cr
\qqq

We finally look for solutions to the eigenvalue problem 
for the Hamiltonians (\ref{tildeR0}) of the form 
\qq\label{asymp}
\Upsilon(u,\un z)=\ee^{\lle\xi,u\rle}F(u,\un z)
\qqq
such that $\xi\in{\gl h}^\NC$ and $F(u/\epsilon, \un z)$ 
is for generic $u$ and $\un z$ holomorphic in $\epsilon\in\NC$,
i. e. we assume 
\qq\label{expan}
F(u/\epsilon, \un z) = F_0(\un z)+ \epsilon F_1(u,\un z) +\dots
\qqq
If solutions of this form exist (later we shall see that this is the 
case) then $F_0$ is an  eigenvector of Hamiltonians,
$R_n=\lim_{u\to\infty}\ee^{-\lle\xi,u\rle}\tilde R_n \ee^{\lle\xi,u\rle},$
and 

\qq\label{Rn}
R_n=\sum\limits_{m\not=n}{t^a_{_{\s(m)}} t^a_{_{\s(n)}}\over
z_{m}-z_{n}}+\xi{_{(n)}}
\qqq

In the case of $SU(2)$ these are the conserved quantities build
by CRS \cite{CRS}. The reduced BCS Hamiltonian is essentially given
by $H=\sum_n z_n R_n$ (\ref{BCS})
and hence it obviously commutes with all $R_n$.

\nsection{Bethe ansatz and horizontal sections.}

We shall address now the question of finding solutions to the
equations that appear in the different limits.
Following \cite{FV} we define for $\beta\in{\gl h}^*$
$$
\omega_\beta(z)=\sigma_{\lle u,\beta \rle} (z) dz +
{i\over 2\pi} \partial_x \sigma_{\lle u,\beta \rle} (z) d\tau
$$
note that $\omega$ is a closed form in $z, \tau$ space.

Now decompose the sum of highest weights in simple roots
(we assume that the decomposition exists):
$$\sum_{s=1}^K\alpha_s=\sum_{n=1}^N\Lambda_n$$
and for any partition $\underline K=(K_1,\cdots,K_N)$
of $K$ in $N$ non-negative integers,
relabel the roots $\underline\alpha=(\alpha_1, \dots, \alpha_K)\equiv
(\alpha_{1,1},\dots,\alpha_{1,{K_1}},\alpha_{2,1},\dots,\alpha_{N,K_N})$.
Introduce a complex variable $y_s$ for any simple root $\alpha_s$
and denote $\underline y=(y_1,\dots,y_K)\equiv(y_{1,1},
\dots,y_{1,{K_1}},y_{2,1},\dots,y_{N,K_N})$.

Set the form
\qq
\begin{array}{l}
\displaystyle \phantom{\Big(}
 \Omega_{\underline K,\underline\alpha}(\tau, u, \underline z,\underline y)
= \omega_{\beta_{1,1}}(y_{1,1}-y_{1,2})\wedge  
 \omega_{\beta_{1,2}}(y_{1,2}-y_{1,3})\wedge\cdots
 \wedge\omega_{\beta_{1,K_1}}(y_{1,K_1}-z_1)\\
\displaystyle
\phantom{\Big(
 \Omega_{\underline K,\underline\alpha}(\tau, u, \underline z,\underline y)
}
\wedge\ 
\cdot\ \cdot\ \cdot\ \cdot\ \cdot\ \cdot\ \cdot\ \cdot\ 
 \cdot\ \cdot\ \cdot\ \cdot\ \cdot\ \cdot\ \cdot\ \cdot\ \cdot\ \cdot\  
 \cdot\ \cdot\ \cdot\ \cdot\ \cdot\ \cdot\ \cdot\ 
\cdot\ \cdot\ \cdot\ \cdot\ \cdot\ \cdot\
\\
\displaystyle
\phantom{\Big(
 \Omega_{\underline K,\underline\alpha}(\tau, u, \underline z,\underline y)
} \wedge\
\omega_{\beta_{N,1}}(y_{N,1}-y_{N,2})\wedge  
 \omega_{\beta_{N,2}}(y_{N,2}-y_{N,3})\wedge\cdots\wedge
 \omega_{\beta_{N,K_N}}(y_{N,K_N}-z_N)\\
\phantom{\Big(
 \Omega_{\underline K,\underline\alpha}(\tau, u, \underline z,\underline y)
}\displaystyle
\otimes_n(e_{-\alpha_{n,1}}\cdots 
e_{-\alpha_{n,K_n}})_{(n)}|\Lambda\rangle
\end{array}
\label{Omega}
\qqq
where $\beta_{n,i}=\sum_{i'=1}^i\alpha_{n,i'}$ and 
$|\Lambda\rangle=\otimes_n|\Lambda_n\rangle$ stands for the tensor 
product of highest weight states of representations $\Lambda_n$.

And finally define
$$\Omega(\tau,u,\underline z,\underline y)=
\sum_{\underline K}
\sum_{\CP\in S_K} 
(-1)^{|\CP|}\Omega_{\underline K,\CP \underline\alpha}
(\tau,u,\underline z,\CP\underline y)
$$

The form $\Omega$ satisfies the equation
\qq\label{crucial}
\partial S\wedge\Omega+d\tau\wedge H_0\Omega+
\sum_n dz_n\wedge H_n\Omega=0
\qqq
where $S$ is the multivalued holomorphic function
\qq
\begin{array}{l}
\displaystyle
S(\tau,\underline z,\underline y)= 
\sum_{n<n'}\langle\Lambda_n,\Lambda_{n'}\rangle
\ln\tilde\vartheta_1(z_n-z_{n'})-
\sum_{n,s}\langle\Lambda_n,\alpha_s\rangle
\ln\tilde\vartheta_1(z_n-y_s) \\
\displaystyle
\phantom{S(\tau,\underline z,\underline y)= }
+\sum_{s<s'}\langle\alpha_s,\alpha_{s'}\rangle
\ln\tilde\vartheta_1(y_s-y_{s'})
\end{array}
\label{ff}
\qqq
with $\tilde\vartheta_1(z)=\vartheta_1(z)/\vartheta_1'(0)$,
and $\partial$ stands for the holomorphic differential
in $\tau,\underline z$ and $\underline y$ variables. 

Equation (\ref{crucial}) was stated in \cite{FV}
and plays the key role in order to get
horizontal sections of the KZB connection 
(\ref{KZB}) from 
integrals of form $\Omega$ in cycles with
coefficients in a local system
determined by $\ee^{{1\over\kappa}S}$. 
See \cite{FV} for more details.

We are interested in a complementary aspect
of eq. (\ref{crucial}) that relates it directly
to the Bethe ansatz solutions of Hamiltonians $H_i$.
Let the function
$G(\tau,u,\underline z,\underline y)$ be given by
$$ G\ d\tau\wedge d^Nz\wedge d^Ky =d \tau\wedge d^Nz\wedge\Omega.$$
Then for any configuration $\underline y$ that satisfies
the Bethe equations
$$\partial_{y_s} S=0, \quad s=1,\dots,K$$
one derives from (\ref{crucial})
\qq
\begin{array}{l}
\displaystyle
H_0\, G = -(\partial_\tau S)\,G\\
\displaystyle
H_n\, G = -(\partial_{z_n} S)\,G
\end{array}
\label{ee}
\qqq
i.e. $G$ is an eigenvector of $H_0$ and $H_n, n=1,\dots,N$ with 
eigenvalues $-\partial_\tau S$ and $-\partial_{z_n} S$, respectively

Here we will need a generalization of these equations
in order to get solutions with the asymptotic behavior
of eq. (\ref{asymp}), that will eventually lead to the 
Richardson equations.
Following \cite{FG3} take $\xi\in{\gl h}^\NC$ and define 
\qq
\tilde\Omega=\ee^{\langle u,\xi\rangle}\Omega,\qquad
\tilde S=S+{1\over 4\pi i}|\xi|^2\tau-
\sum_n\langle\xi,\Lambda_n\rangle z_n+
\sum_s\langle \xi,\alpha_s\rangle y_s,
\qqq
from (\ref{crucial}) one can derive a similar relation for the new 
objects
\qq\label{cructilde}
\partial \tilde S\wedge\tilde\Omega+d\tau\wedge H_0\tilde\Omega+
\sum_n dz_n\wedge H_n\tilde\Omega=0
\qqq
and then one has eigenvectors
\qq
\begin{array}{l}
\displaystyle
H_0 \ee^{\langle u,\xi\rangle} G = -({1\over4\pi i}|\xi|^2 + \partial\tau S)
\ee^{\langle u,\xi\rangle} G
\\
\displaystyle
H_n \ee^{\langle u,\xi\rangle} G = 
(\langle\xi,\Lambda_n\rangle-\partial_{z_n} S)
\ee^{\langle u,\xi\rangle} G
\end{array}
\label{kk}
\qqq
provided $\underline y$ is chosen so that
$$
\sum_{s'\not=s}\langle\alpha_{s'},\alpha_s\rangle\rho(y_{s'}-y_s)
-\sum_n\langle\Lambda_n,\alpha_s\rangle\rho(z_n-y_s)= 
\langle\xi,\alpha_s\rangle
$$

In order to obtain eigenvectors for the Hamiltonians of 
eqs. (\ref{tildeR0}) we redefine
$\xi\mapsto{\xi/\zeta}$ and 
take, as in the previous section, the two consecutive limits
$\tau\to i\infty$ and $\underline z, u\to 0$, i.e. $\zeta\to 0$ 
in G.

After this limiting procedure $G\to F$ where 
\qq\label{F}
F(u,\underline z,\underline y)=\sum_{\underline K}\sum_{\CP\in S_K} 
F_{\underline K,\CP\underline\alpha}
\qqq
with
\qq
\begin{array}{l}
\displaystyle \phantom{\Big(}
F_{\underline K,\underline\alpha}(u,\underline z,\underline y)
= v_{\beta_{1,1}}(y_{1,1}-y_{1,2})  
  v_{\beta_{1,2}}(y_{1,2}-y_{1,3})\cdots
  v_{\beta_{1,K_1}}(y_{1,K_1}-z_1)\\ \displaystyle
\phantom{\Big( F_{\underline K,\underline\alpha}(u,\underline z,\underline y)
= }\times\ \cdot\ \cdot\ \cdot\ \cdot\ \cdot\ \cdot\ \cdot\ 
\cdot\ \cdot\ \cdot\ \cdot\ \cdot\ \cdot\ \cdot\ \cdot\ 
\cdot\ \cdot\ \cdot\ \cdot\ \cdot\ \cdot\ \cdot\ \cdot\ 
\cdot\ \cdot\ \cdot\ \cdot\ \cdot\ \cdot \\
\displaystyle
\phantom{\Big( F_{\underline K,\underline\alpha}(u,\underline z,\underline y)
= }
 \times\ v_{\beta_{N,1}}(y_{N,1}-y_{N,2})  
  v_{\beta_{N,2}}(y_{N,2}-y_{N,3})\cdots
  v_{\beta_{N,K_N}}(y_{N,K_N}-z_N) \\ \displaystyle
\phantom{\Big(  F_{\underline K,\underline\alpha}(u,\underline z,\underline y)
= }
\otimes_n(e_{-\alpha_{n,1}}\cdots 
e_{-\alpha_{n,K_n}})_{(n)}|\Lambda\rangle
\end{array}
\label{FK}
\qqq
where 
$$v_\beta(z)={1\over z}-{1\over \langle u, \beta \rangle }$$
and the rest of the notation is as in eq. (\ref{Omega}).
 
One  then satisfies the eigenvalue equations:
\qq
\begin{array}{l}
\displaystyle
\tilde R_0 \ee^{\langle u,\xi\rangle} F 
= -{1\over 4\pi i}|\xi|^2 
\ee^{\langle u,\xi\rangle} F\\ \displaystyle
\tilde R_n \ee^{\langle u,\xi\rangle} F 
=(\langle\xi,\Lambda_n\rangle-\partial_{z_n}T)
\ee^{\langle u,\xi\rangle} F
\end{array}
\label{jj}
\qqq 
which hold for 
\qq
\begin{array}{l}
\displaystyle
T(\underline z,\underline y)=
\sum_{n<n'}\langle\Lambda_n,\Lambda_{n'}\rangle
\ln(z_n-z_{n'})-
\sum_{n,s}\langle\Lambda_n,\alpha_s\rangle
\ln(z_n-y_s)\\ \displaystyle
\phantom{T(\tau,\underline z,\underline y)=}
+\sum_{s<s'}\langle\alpha_s,\alpha_{s'}\rangle
\ln(y_s-y_{s'})
\end{array}
\label{gg}
\qqq
provided $\un y$ is chosen so that it fulfills the Bethe equations
$$
\partial_{y_s} T+\langle\xi,\alpha_s\rangle=0\qquad s=1,\dots,K
$$
or explicitly
\qq\label{betheeq}
\sum_{s'\not=s}{\langle\alpha_{s'},\alpha_s\rangle\over y_{s'}-y_s}
-\sum_n{\langle\Lambda_n,\alpha_s\rangle\over z_n-y_s}= 
\langle\xi,\alpha_s\rangle \qquad s=1,\dots,K.
\qqq

Note that $F(u/\epsilon,\underline z,\underline y)$ is 
for generic $\underline z, \underline y$ and $u$ a polynomial
in $\epsilon$ thus it admits the expansion of eq. (\ref{expan})
and 
$$F_0(\un z,\un y)=\lim_{\epsilon\to0}F(u/\epsilon,\underline z,\underline y)$$
is an eigenvector of Hamiltonians $R_n$ provided  
$\underline y$ is a solution of (\ref{betheeq}),
whose corresponding  eigenvalue $\mu_n$ 
is given by $\langle\xi,\Lambda_n\rangle-\partial_{z_n}T$, i. e.
\qq\label{mu}
\mu_n=
\langle\xi,\Lambda_n\rangle
+\sum_{n'\not=n}{\langle\Lambda_{n'},\Lambda_n\rangle\over z_{n'}-z_n}
-\sum_s{\langle\alpha_{s},\Lambda_n\rangle\over y_{s}-z_n}
\qqq

In our construction we only considered the subspace
$V^0$ of vectors in the zero total weight subspace, this was imposed
from the very beginning, for only when restricted
to this subspace the Hamiltonians $H_n$ commute, defining  
an integrable system. 
However after the limiting procedure we arrive at Hamiltonians
$R_n$ and these have the property of commuting in the whole
representation space. We should then produce solutions
of arbitrary weight.

Suppose we search for an eigenvector of the 
Hamiltonians $R_n$ of total weight $\lambda_0$.
In order to get it we introduce a new insertion at
$z_0$ with representation of highest weight $\Lambda_0$
and vector space $V_{\Lambda_0}$.
Now we decompose the sum of highest weights  
$$\sum_{s=1}^L \alpha_s =\sum_{n=1}^N\Lambda_n + \Lambda_0$$
with $\Lambda_0$ and the last $L-K$ roots chosen to verify
$$\lambda_0=\sum_{s=K+1}^L \alpha_s -\Lambda_0.$$
We reparametrize the corresponding screening variables
$$(y_1,\dots,y_L)\equiv
(y_1,\dots,y_K,z_0+\gamma_1,\dots, z_0+\gamma_{L-K})$$
and send $z_0$ to $\infty$ while keeping $y_s$ and $\gamma_l$ finite.
In the limit, the old Hamiltonians $R_n$ decouple from the new insertion,
the Bethe equations (\ref{betheeq}) for $s=1,\dots,K$ are unchanged
(the new terms coming from the new insertion and screening variables 
vanish in the limit) and the eigenvectors factorize in $F_0\otimes f$
where $f\in V_{\Lambda_0}$ has weight $-\lambda_0$. As the total weight is 
zero $F_0$ is an eigenvector of Hamiltonians $R_n$ (provided the
Bethe equations are fulfilled) of weight $\lambda_0$.
In practice we can forget about the new insertion and 
screening variables and simply solve the Bethe equations (\ref{betheeq})
and use the Bethe ansatz solution (\ref{F},\ref{FK}) but with a new set
of simple roots so that
\qq\label{weights}
\sum_{s=1}^K \alpha_s =\sum_{n=1}^N \Lambda_n-\lambda_0,
\qqq
and accordingly a different number of screening variables $y_s$.

This ends our main results which show the deep mathematical
structure underlying the connection of 
BCS-like Richardson models with Chern-Simons theory.

\nsection{Conclusions}

In this paper we have shown how the BCS model,
its exact solution, and integrability is
related to the Chern-Simons theory. This has been
done in two ways. First by adding a perturbation
to the CS action on the plane by means of a 
one-dimensional coloured charge distribution, 
and second by taking an appropriated limit
of the CS model on a torus 
with twisted boundary conditions
for the WZW gauge fields. 
The construction is valid for any simple, simply
connected compact Lie group $G$. The case of BCS
corresponding to the choice $G = SU(2)$.

We find quite remarkable that the Chern-Simons
theory not only gives effective descriptions
of the fractional quantum Hall effect and 
the anyon superconductivity, 
but also of the standard BCS superconductivity. 
Though in the later case the formalism
works, not in real space, but in the
complex energy space.
The space dimensionality enters only through
the representations of the Lie
group associated to every energy level,
which for BCS is simply the level degeneracy.

This work adds one more member to the
series: BCS/Gaudin models $\rightarrow $
Integrable Vertex Models $\rightarrow $
WZW models $\rightarrow $ Chern-Simons theory, 
suggesting new research directions. One
is the possible existence of integrable models
associated to higher genus Riemann surfaces.
This is an interesting but difficult problem, see \cite{hg} 
for a study of the genus 2 case.

Another question comes from 
the free field realization of the 
$SU(2)$ WZW model. This model has two screening
operators $J_\pm$, where only one of them, i.e.
$J_+$, is required for the computation of correlators
of the WZW primary fields with integer spins,
while the second one is needed for primary fields
with fractional spin \cite{BO,D1,D2}. In the 
limit where $k+2 \rightarrow 0$, the positions
of the screening operators $J_+(u)$ get frozen
and satisfy the Richardson/Gaudin equations
\cite{S}. Thus $J_+$  can be
regarded as a Cooper pair creation operator. 
The question is: what is the role, if any,
of the second screening operator $J_-$
and the
associated primary fields with fractional spin?
A conjecture is that they are associated to vortex
operators in the BCS model since they produced
additional zeros in the wave function. This
possibility is currently under investigation.

\vspace{1cm}
{\bf Acknowledgements} 
G.S. would like to thanks E.H. Kim and J. Dukelsky for discussions
and the Department of Theoretical Physics of Zaragoza University
for  hospitality. 
We would like to thanks the Benasque Center of 
Physics where this work was completed. 
This work has been supported by the Spanish MCyT grants 
FPA2000-1252, BFM2000-1320-C02-01.

\nappendix{}
\vskip 0.5cm

The aim of this  appendix is to give a more explicit
expression of the basic equations (\ref{betheeq},\ref{mu}) for their
application to several problems in Condensed Matter
and Nuclear Physics. To this end we shall exploit the
structure of the simple Lie groups, together with their 
representation theory \cite{dF}. 

The problem that is posed is the following: given a set of 
$N$ complex variables  $\{ z_n \}_{n=1}^N$ together
with a set of highest weights (h.w.) $\{ \Lambda_n \}_{n=1}^N$
of a simple Lie group $G$ and 
a generic element $\xi$ belonging to the Cartan subalgebra
${\gl h}^{\NC}$ of $G$, find the solutions of 
``semiclassical Bethe equations'' (\ref{betheeq}) for a 
set of $K$ unknowns $\{ y_s \}_{s=1}^K$, which
are associated to a set of simple roots $\{ \alpha_s \}_{s=1}^K$.
Moreover the sets of h.w. and simple roots are related 
by the eq. (\ref{weights}) where $\lambda_0$ is the total weight
of the eigenvector we look for. 
Once the solutions of (\ref{betheeq}) are found 
then eqs.(\ref{mu}) give the eigenvalues $\mu_n$ of the 
``perturbed'' Gaudin operators $R_n$ defined
in eqs. (\ref{Rn}). 

A simple Lie Group $G$, with 
rank $r$, is characterized by $r$ simple roots 
$\{ \balpha_a \}_{a=1}^r$. It is convenient
to describe  the set  $\{ \alpha_s \}_{s=1}^K$ 
by the numbers $\{N_a \}_{a=1}^r$ of roots
of the same type $\balpha_a$. Then the label
$s$ amounts to the pair of labels $(a,i)$ with 
$a=1, \dots,r$ and $i=1, \dots, N_a$, such that
$K = \sum_{a=1}^r \; N_a$. Correspondingly  we
shall make the change of variables: 
$y_s = y_{(a,i)} \equiv E^a_i$. 
With these definitions eqs.(\ref{betheeq}) become
\begin{eqnarray}
\sum_{b=1}^r \sum_{j=1}^{N_b} \; 
\frac{ \langle \balpha_b, \balpha_a \rangle }{E^b_j - E^a_i}
- \sum_{n=1}^N 
 \frac{ \langle \Lambda_n, \balpha_a \rangle }{z_n - E^a_i}
= \langle \xi, \balpha_a \rangle
\label{a1}
\end{eqnarray}
\noindent where $a=1, \dots,r$ and $i=1, \dots, N_a $. Of course
the term $i=j, a=b$ must be excluded in (\ref{a1}). 
Similarly eqs. (\ref{mu}) read,
\begin{eqnarray}
\mu_n = \langle \xi, \Lambda_n \rangle + 
\sum_{n' \neq n} \frac{ \langle \Lambda_{n'}, \Lambda_n \rangle }{
z_{n'} - z_n} - \sum_{a=1}^r \sum_{i=1}^{N_a} 
\frac{ \langle \balpha_a, \Lambda_n \rangle}{ E^a_i - z_n}
\label{a2}
\end{eqnarray}

The various scalar products $ \langle \cdot, \cdot \rangle$ 
appearing in these eqs. are well known in the theory
of Lie algebras \cite{dF}. First of all, the scalar 
product  of simple roots 
$\langle \balpha_a, \balpha_b \rangle$ is directly
related to the Cartan matrix
$A_{a  b}$, which is defined as
\begin{eqnarray}
A_{a  b} = \langle \balpha_a, \balpha_b^{\vee} \rangle 
\label{a3}
\end{eqnarray}
\noindent where $ \balpha_a^{\vee}$ is the coroot 
of $ \balpha_a$ given by 
\begin{eqnarray}
\balpha_a^{\vee} = \frac{ 2 \balpha_a}{ | \balpha_a|^2 } 
\label{a4}
\end{eqnarray}

For simply laced Lie groups (i.e. the ones in the ADE series) 
and in the normalization we have adopted, 
one has $| \balpha_a|^2=2, \; \forall a$,
so that roots and coroots are equal, and the Cartan
matrix is symmetric. 

Next we have to compute the scalar products involving
the h.w.'s $\Lambda_n$.  Every weight $\lambda$ is 
characterized by its expansion in the basis of the fundamental
weight vectors $\{ \omega_a \}_{a=1}^r $, i.e. 
\begin{eqnarray}
\lambda = \sum_{a=1}^r \; \lambda^a \; \omega_a 
\label{a5}
\end{eqnarray}
\noindent where $\{ \lambda^a \}_{a=1}^r $ are
integers called Dynkin labels (they are non negative for
h.w.v). The basic property of the fundamental weights is 
\begin{eqnarray}
\langle \omega_a, \balpha_b^\vee \rangle = \delta_{ab}
\label{a6}
\end{eqnarray}
\noindent 
Using eqs(\ref{a4},\ref{a5},\ref{a6}) one can deduce
\begin{eqnarray}
\langle \lambda, \balpha_a \rangle = \frac{ |\balpha_a|^2}{2} 
\lambda^a 
 \label{a7}
\end{eqnarray}
\noindent 
The scalar product of two weights $\lambda_1$ and  $\lambda_2$
is given by 
\begin{eqnarray}
\langle \lambda_1, \lambda_2 \rangle = \sum_{a,b=1}^r \lambda_1^a 
F_{a b} \lambda_2^b \equiv \lambda_1 \cdot F \cdot \lambda_2
\label{a8}
\end{eqnarray}
\noindent where $F_{a b}$ is called the quadratic form matrix
which is defined as 
\begin{eqnarray}
F_{a b} = \langle \omega_a , \omega_b \rangle = (A^{-1})_{a,b}
\frac{ |\balpha_b|^2}{2}
\label{a9}
\end{eqnarray}
\noindent 
The last equality in eq.(\ref{a9})  follows from the relation between
simple roots and fundamental weights,
\begin{eqnarray}
\balpha_a = \sum_{b=1}^r A_{a b } \omega_b 
\label{a10}
\end{eqnarray}
\noindent 
The element $\xi \in {\gl h}^{\NC}$ will also be expanded 
in the basis of fundamental weights, 
\begin{eqnarray}
\xi = \sum_{a=1}^r \xi^a  \omega_a
\label{a11}
\end{eqnarray}
\noindent Then using eqs. (\ref{a7}) and (\ref{a8}) we get,
\begin{eqnarray}
\begin{array}{l}
\displaystyle
\langle \xi, \balpha_a \rangle =  \frac{ |\balpha_a|^2}{2} \; \xi^a   \\
\langle \xi, \lambda \rangle =    \xi \cdot F \cdot \lambda,
\end{array}
\label{a12}
\end{eqnarray}
\noindent 
Using the formulae introduced above
we can finally write the Bethe eqs.(\ref{a1}) as
\begin{eqnarray}
\sum_{b=1}^r \sum_{j=1}^{N_b} \; 
\frac{ A_{b a} }{E^b_j - E^a_i}
- \sum_{n=1}^N 
 \frac{ \Lambda_n^a }{z_n - E^a_i}   = \xi^a
\label{a13}
\end{eqnarray}
\noindent and the eigenvalues (\ref{a2}) as 
\begin{eqnarray}
\mu_n =   \xi \cdot F \cdot \Lambda_n
 + \sum_{n' \neq n} \frac{ \Lambda_{n'} \cdot F \cdot \Lambda_n }{
z_{n'} - z_n} - \sum_{a=1}^r \sum_{i=1}^{N_a} 
\frac{ \Lambda_n^a |\balpha_a|^2 / 2}{ E^a_i - z_n}
\label{a14}
\end{eqnarray}
\noindent where $\Lambda_n^a $ are the Dynkin labels of the
h.w. $\Lambda_n$. In reference \cite{dF} 
can be found the explicit
expressions of the Cartan and quadratic form matrices, so the remaining
data to be fixed are the representations associated to the sites
as well as the weight $\lambda_0$ (see below for examples).

{\bf Example 1: $G=SU(2)$}

$SU(2)$ is the unique simple Lie group with rank 1 (i.e.
$r=1$), with a simple root $\balpha_1 = \sqrt{2}$ and a fundamental
weight $\omega_1 = \balpha_1/2 = 1/\sqrt{2}$. The Cartan
and  the quadratic form ``matrices'' are 
given by   $A_{11} = 2$ and $F_{11} = 1/2$. The Dynkin 
label of a h.w.  $\Lambda$ is twice the spin, i.e. 
$\Lambda^1= 2 s, \; (s=0, 1/2, 1, \dots)$. With the 
notation $N_1=M$ (M being the number
of Cooper pairs) and $\xi^1 = -1/g$, ($g$ being the BCS coupling constant),
eqs.(\ref{a13}) yield  the Richardson eqs. (\ref{6})
\begin{eqnarray}
\sum_{j \neq i }^{M} \; 
\frac{ 2 }{E_j - E_i}
- \sum_{n=1}^N 
 \frac{ 2 s_n }{z_n - E_i} + \frac{1}{g} = 0
\label{a15}
\end{eqnarray}
\noindent associated to the BCS Hamiltonian,  
while eqs.(\ref{a14}) yield the eigenvalues
of the CRS conserved quantities \cite{CRS,S,DS}. 
\begin{eqnarray}
\mu_n = - \frac{s_n}{g}  
 + \sum_{n' \neq n} \frac{ 2 s_{n'} s_n  }{
z_{n'} - z_n} -  \sum_{i=1}^{M} 
\frac{ 2 s_n }{ E_i - z_n}
\label{a16}
\end{eqnarray}

In the BCS case, $N$ is the number of energy
levels which are occupied by electrons pairs,  
$z_n$ is twice the single particle
energy (i.e. $z_n = 2 \epsilon_n$), $2 s_n$ is the maximal number
of pairs that can occupy that level. 

Denoting by  $s_0$ the spin  of the weight $\lambda_0$
appearing in eq.(\ref{weights})  one gets
\begin{eqnarray}
s_0 = \sum_{n=1}^N s_n  - M
\label{a17}
\end{eqnarray}
\noindent In the case where $s_n= 1/2, \; \forall n$,
the half-filling constraint (i.e. as 
many electrons as energy levels) is given by 
$N = 2 M$, which from (\ref{a17}) corresponds to 
$s_0=0$.

Before we consider more general groups, we 
shall discuss in more detail eq.(\ref{weights}) regarding the relation
between the number of roots $N_a$ and the weights $\Lambda_n$
and $\lambda_0$. First of all this eq. can be written as
\begin{eqnarray}
\sum_{a=1}^r N_a \balpha_a = \sum_{n=1}^N \Lambda_n - \lambda_0
\label{a18}
\end{eqnarray}

The problem is to find the numbers $N_a$ given  
$N, \{ \Lambda_n \}_{n=1}^N$ and $\lambda_0$. Actually
there may not exists a solution of this eq. with
all $N_a$ non negative integers.

A particular case where we know there must 
exist a solution is for all the $\Lambda_n's$ 
the adjoint irrep and $\lambda_0=0$. The h.w. of the adjoint 
representation is given by the highest root $\theta$ 
\begin{eqnarray}
\theta = \sum_{a=1}^r d_a \balpha_a 
\label{a19}
\end{eqnarray}
\noindent where $d_a$ are positive integers 
called the marks (we use a different notation from reference.
 \cite{dF}). For $SU(n)$ groups $d_a= 1, \; \forall a$. 
Setting $\Lambda_n = \theta, \; \forall n $ and $\lambda_0= 0$
in (\ref{a18}) the  solution is given by
\begin{eqnarray}
N_a = d_a \; N 
\label{a20}
\end{eqnarray}

Let us now consider the more general case. 
Using eq.(\ref{a10}) in (\ref{a18}) we get
\begin{eqnarray}
\sum_{b=1}^r N_b A_{b a} = \sum_{n=1}^N \Lambda^a_n - \lambda_0^a
\label{a21}
\end{eqnarray}
\noindent Multiplying by the inverse of the Cartan matrix
and using eq.(\ref{a9}) we finally get the desired expression:
\begin{eqnarray}
N_a = \frac{2}{|\balpha_a|^2} \sum_{b=1}^r F_{b a}
\left( \sum_n \Lambda_n^b - \lambda_0^b \right) 
\label{a22}
\end{eqnarray}
\noindent which obviously may not yield integers values
for $N_a$.

{\bf Example 2: $G=SU(r+1)$}

The Lie group $A_r \equiv  SU(r+1)$ has rank $r$.
We shall choose $N$ insertions in the fundamental representation,  
whose Dynkin labels are given by $\Lambda = (1,0, \dots,0)$. From 
eq.(\ref{a21}) we derive then
\begin{eqnarray}
\begin{array}{ll}
\lambda_0^1 = & N - 2 N_1 + N_2 \\
\lambda_0^2 = & -2 N_2 + N_1 + N_3 \\
\vdots & \vdots \\
\lambda_0^r = & - 2 N_r + N_{r-1}.
\end{array}
\label{a23}
\end{eqnarray}

If we choose $\lambda_0=0$
then the number of roots $N_a$ follows from eq.(\ref{a22})
\begin{eqnarray}
N_a = N\ \frac{r + 1 -a}{r+1}
\label{a24}
\end{eqnarray}
\noindent Hence $N$ must be a multiple of $r+1$. 
This reflects the fact that the tensor product
of $r+1$ copies of the fundamental irrep contains the identity.

The Bethe eqs. in the case of $SU(3)$ and arbitrary irreps
is given by 
\qq
\begin{array}{l}
\displaystyle
\sum_{j=1}^{N_1} \; 
\frac{ 2 }{E^1_j - E^1_i}- 
\sum_{j=1}^{N_2} \; 
\frac{ 1 }{E^2_j - E^1_i}
- \sum_{n=1}^N 
 \frac{\Lambda_n^1}{z_n - E^1_i}  = \xi^1 
\\
\displaystyle
\sum_{j=1}^{N_2} \; 
\frac{ 2 }{E^2_j - E^2_i}- 
\sum_{j=1}^{N_1} \; 
\frac{ 1 }{E^1_j - E^2_i}
- \sum_{n=1}^N 
 \frac{\Lambda_n^2}{z_n - E^2_i}
 = \xi^2 \\
\end{array}
\label{a25}
\qqq

Other representations and groups can be worked
out similarly.

\end{document}